# An Irregular Two-Sizes Square Tiling Method for the Design of Isophoric Phased Arrays

Paolo Rocca, *Senior Member, IEEE*, Nicola Anselmi, *Member, IEEE*, Alessandro Polo, *Member, IEEE*, and A. Massa, *Fellow, IEEE*

*Abstract*—The design of isophoric phased arrays composed of two-sized square-shaped tiles that fully cover rectangular apertures is dealt with. The number and the positions of the tiles within the array aperture are optimized to fit desired specifications on the power pattern features. Towards this end, starting from the derivation of theoretical conditions for the complete tileability of the aperture, an ad-hoc coding of the admissible arrangements, which implies a drastic reduction of the cardinality of the solution space, and their compact representation with a graph are exploited to profitably apply an effective optimizer based on an integer-coded Genetic Algorithm. A set of representative numerical examples, concerned with state-of-the-art benchmark problems, is reported and discussed to give some insights on the effectiveness of both the proposed tiled architectures and the synthesis strategy.

## I. INTRODUCTION

NEW space technologies are pushing the research towards a new generation of antenna systems for space applications to fit more and more stringent requirements in terms of pattern performance, power consumption, and geometrical constraints [1]. For instance, let us consider the CubeSat missions [1]-[5] based on swarms of cheap miniaturized satellites for communications and sensing as well as recent investigations on space-based infrastructures for Internet-of-Space (*IoS*) applications [6]. All these applicative scenarios need

Manuscript received on October XX, 2019
This work was been partially supported by the Italian Ministry of Education, University, and Research within the Program "Smart cities and communities and Social Innovation" (CUP: E44G14000060008) for the Project "WATERTECH - Smart Community per lo Sviluppo e l'Applicazione di Tecnologie di Monitoraggio Innovative per le Reti di Distribuzione Idrica negli usi idropotabili ed agricoli" (Grant no. SCN_00489) and within the Program PRIN 2017 (CUP: E64I19002530001) for the Project "CYBER-PHYSICAL ELECTROMAGNETIC VISION: Context-Aware Electromagnetic Sensing and Smart Reaction (EMvisioning)" (Grant no. 2017HZJXSZ).

P. Rocca, N. Anselmi, A. Polo, and A. Massa are with the ELEDIA@UniTN (DISI - University of Trento), Via Sommarive 9, 38123 Trento - Italy (e-mail: {paolo.rocca, nicola.anselmi.1, alessandro.polo.1, andrea.massa}@unitn.it).

P. Rocca, is also with the ELEDIA Research Center (ELEDIA@XIDIAN - Xidian University),P.O. Box 191, No.2 South Tabai Road, 710071 Xi'an, Shaanxi Province - China (e-mail: paolo.rocca@xidian.edu.cn).

A. Massa is also with the ELEDIA Research Center (ELEDIA@L2S - UMR 8506), 3 rue Joliot Curie, 91192 Gif-sur-Yvette - France (e-mail: andrea.massa@l2s.centralesupelec.fr).

A. Massa is also with the ELEDIA Research Center (ELEDIA@UESTC - University of Electronic Science and Technology of China), School of Electronic Engineering, 611731 Chengdu - China (e-mail: andrea.massa@uestc.edu.cn).

A. Massa is also with the ELEDIA Research Center (ELEDIA@TSINGHUA - Tsinghua University), 30 Shuangqing Rd, 100084 Haidian, Beijing - China (e-mail: andrea.massa@tsinghua.edu.cn).

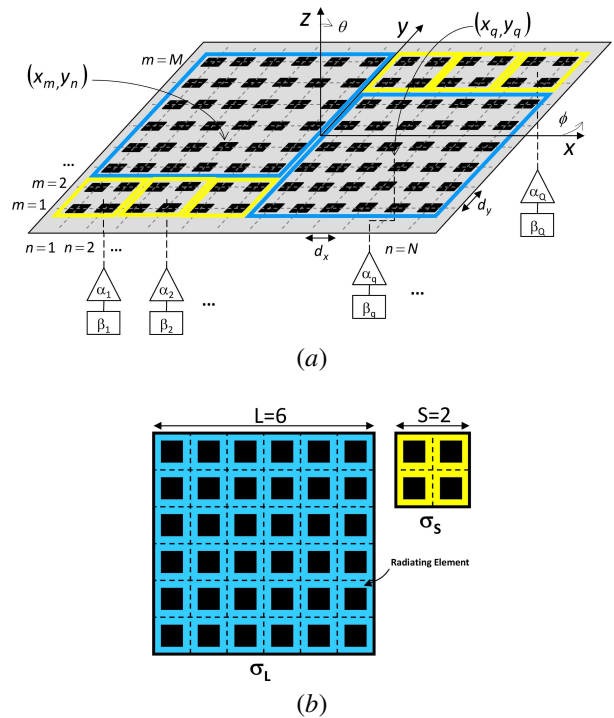

Figure 1.   Sketch of (*a*) the *MSTA* architecture and (*b*) an example of $\sigma_S$ and $\sigma_L$ tile arrangement when $S = 2$ and $L = 6$.

high-efficiency, low-profile, and low-cost antennas to be easily packaged into small payloads and lunched in the space with reduced costs. Moreover, current space programs have shown an increasing interest in array solutions for sensing [7][8] and *IoS* [9] because of their advanced and attractive functionalities. However, conventional (fully-populated) phased arrays (*FPAs*) imply the fabrication of complex beamforming networks (*BFNs*) with many transmission/receive modules (*TRMs*), generally one for each element of the array, thus resulting unaffordable for space applications. The need of lowering the fabrication costs of phased arrays has induced the antenna research community to investigate alternative architectures to fully-populated arrangements for reducing the number of *TRMs* in the *BFN*, as they represent the main source of cost [10], while still keeping the positive features of phased array antennas. Consequently, unconventional compromise arrays have been proposed [11] including sparse [12]-[19], thinned [20][21], and clustered architectures [22]-[31]. Among these latter, tiled layouts recently attracted a large interest being a promising architectural solution for a future mass deployment of phased arrays. Thanks to their modularity (the array being composed of one or few types of tiles), the production and the maintenance costs can be significantly reduced as well as the *BFN* turns out to be simplified since each tile is controlled by a single *TRM*. However, designing a tiled array is not a trivial task and it is more and more difficult when dealing with irregular clusters to mitigate the presence of undesired grating lobes and to reduce high quantization lobes [25][29]. Moreover, a complete coverage of the antenna aperture is mandatory when challenging directivity requirements must be satisfied. From a functional point of view, both the exact tiling of a finite region - given a limited set of tile shapes - and the generation of the set of admissible and complete clusterings are challenging combinatorial tasks. Fortunately, these topics have been addressed in several theoretical contributions [32]-[37] in the mathematical literature and useful coverage theorems, along with exhaustive generation algorithms, have been derived. Recently, such a theoretical/algorithmic background has been fruitfully exploited and profitably customized to the irregular tiling of rectangular and hexagonal array apertures with domino and diamond shapes [29]-[31], respectively.

Besides previous challenges in designing tiled arrays, another fundamental need of standard applications of space antennas is the maximum efficiency of the beam-forming amplifiers since the energy budget is extremely tight. Therefore, isophoric arrays (i.e., arrays of elements controlled by amplifiers working at the maximum efficiency) organized in sparse [13][14][16]-[18] or clustered [22][15] configurations have been proposed in the last years by recurring to a density- [13][16]-[18] or an element-size [14][15] tapering to synthesize desired pattern shapes, as well.

Dealing with isophoric tiled arrays, this work is concerned with an innovative synthesis method that guarantees the full coverage of the antenna aperture by means of an irregular placement of uniformly-fed tiles having two square sizes (i.e., each square size clusters a different number of elementary radiators). Thanks to the exploitation of different tile dimensions and the arising amplitude *tile-size* tapering of the aperture illumination, an advanced control of the beam-pattern features is yielded. To the best of the authors' knowledge, the main novelties of this research activity lie in (*i*) the exploitation of suitable mathematical theorems to solve the problem of fully covering rectangular apertures with two square tiles of different sizes; (*ii*) the design of an array architecture composed of two square isophoric sub-arrays with different number of elements, denoted as *Multi-Square Tiled Array* (*MSTA*), to yield an advanced control of the beam-pattern features through an amplitude *tile-size* tapering of the single-element excitations; (*iii*) the definition of an ad-hoc binary coding of all the possible rows composing the planar array and the representation of the whole set of complete aperture tilings as a graph where each node corresponds to one admissible row of the array and each different arrangement of the tiles is mapped into a path of connected nodes; (*iv*) the development of an innovative Integer-coded *GA* (*IGA*) for an efficient exploration of the solution space/graph to effectively deal with large arrays, as well.

The rest of the paper is organized as follows. The mathematical formulation of the synthesis of tiled arrays is summarized in Sect. 2, while Section 3 details the proposed tiling method by including the theoretical conditions for the complete tileability of the aperture (Sect. 3.1), the coding of the admissible clustered configurations as well as the mapping of these latter in a suitable minimum-redundancy graph-based representation (Sect. 3.2), and the optimization approach for the definition of the optimal layout fitting the user-defined pattern requirements (Sect. 3.3). Representative examples from a wide numerical assessment are reported in Sect. 4 where test cases, also concerned with full-wave modeled realistic radiating elements, are discussed. Eventually, some conclusions and final remarks are drawn (Sect. 5).

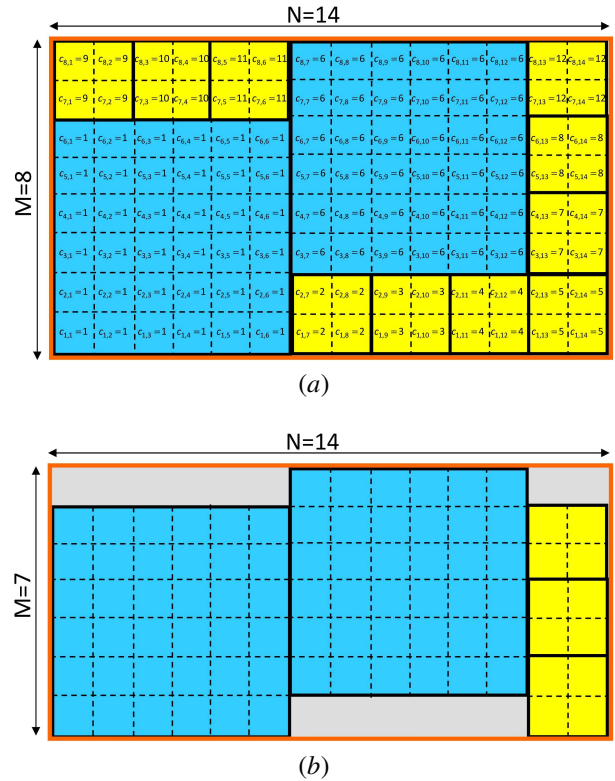

Figure 2. *Coverage Theorem* ($S = 2$, $L = 6$) - Example of (*a*) a tileable aperture ($M = 8$, $N = 14$) and of (*b*) a non-tileable aperture ($M = 7$, $N = 14$).

## II. Mathematical Formulation

Let us consider a planar phased array composed of $M \times N$ elementary radiators placed on a regular rectangular lattice with inter-element distances $d_x$ and $d_y$ along the $x$ and the $y$ axis, respectively [Fig. 1(*a*)]. The antenna aperture $R$ is completely filled in with $Q$ square tiles having two different sizes, but each one has a single input/output port. The two types of tile, $\sigma_S$ and $\sigma_L$, contain $\gamma_S$ ($\gamma_S \triangleq S \times S$) and ($\gamma_L \triangleq L \times L$) radiating elements [$L = 6$, $S = 2$ - Fig. 1(*b*)], respectively, $L$ being an integer multiple of $S$ (i.e., $\hat{L} \triangleq \frac{L}{S}$ is an integer number and $\hat{L} \geq 2$) to set an additional degree of modularity that further simplifies the array manufacturing (i.e., the larger tile $\sigma_L$ can be generated by assembling a set of $\sigma_S$ tiles). Moreover, the $Q$ sub-arrays are assumed to have uniform amplitudes (i.e., $\boldsymbol{\alpha} = \mathbf{1}$, $\mathbf{1} = \{\alpha_q \triangleq 1; q = 1, ..., Q\}$) to avoid any amplitude tapering [18]. The pattern control is yielded by means of the phases of the $Q$ clusters, $\boldsymbol{\beta} = \{\beta_q; q = 1, ..., Q\}$, along with the two-level equivalent amplitude distribution within the aperture. This latter is due to the different number of array elements clustered in the $\sigma_S$ and $\sigma_L$ tiles. Indeed, if each $q$-th ($q = 1, ..., Q$) tile is fed by the same amount of power (i.e., $\alpha_q \triangleq 1$), then the equivalent element-level amplitude excitations, $\{\alpha_{m,n}^{(q)}; m = 1, ..., M; n = 1, ..., N;$ $q = 1, ..., Q\}$, turn out to be equal to

$$\alpha_{m,n}^{(q)} = \begin{cases} \frac{1}{\sqrt{\gamma S}} & if \ (x_n, y_m) \in \sigma_S \\ \frac{1}{\sqrt{\gamma L}} & if \ (x_n, y_m) \in \sigma_L \end{cases};$$
$$m = 1, ..., M; n = 1, ..., N \qquad (1)$$

being $\alpha_q = \sum_{m=1}^{M} \sum_{n=1}^{N} \alpha_{m,n}^{(q)} \delta_{c_{m,n},q}$, while the far-field array radiation can be mathematically expressed as

$$\mathbf{E}(\theta, \phi; \mathbf{c}, \boldsymbol{\beta}) = \sum_{m=1}^{M} \sum_{n=1}^{N} \mathbf{e}_{m,n}(\theta, \phi) \Lambda_{m,n}(\theta, \phi) \times I_{m,n}(\mathbf{c}) \qquad (2)$$

where $\mathbf{e}_{m,n}(\theta, \phi)$ is the active element pattern of the $(m, n)$-th ($m = 1, ..., M$, $n = 1, ..., N$) radiator located at $(x_n, y_m)$, $\Lambda_{m,n}(\theta, \phi) = e^{j\frac{2\pi}{\lambda} \sin\theta(x_n \cos\phi + y_m \sin\phi)}$ is the corresponding steering vector, $\lambda$ being the wavelength, and $I_{m,n}(\mathbf{c})$ is the related weighting term $[I_{m,n}(\mathbf{c}) \triangleq \sum_{q=1}^{Q} \alpha_{m,n}^{(q)} \delta_{c_{m,n},q} e^{j\beta_q}]$, while $(\theta, \phi)$ are the angular directions ($\theta \in [0 : 90]$ [deg], $\phi \in [0 : 360]$ [deg]). Moreover, $\mathbf{c}$ is the tiling vector ($\mathbf{c} \triangleq \{c_{m,n} \in [1 : Q]; m = 1, ..., M, n = 1, ..., N\}$) whose $(m, n)$-th entry, $c_{m,n}$, indicates the membership of the $(m, n)$-th ($m = 1, ..., M; n = 1, ..., N$) elementary radiator to the $q$-th ($q = 1, ..., Q$) tile, $\delta_{c_{m,n},q}$ being the Kronecker delta function equal to $\delta_{c_{m,n},q} = 1$ if $c_{m,n} = q$ and $\delta_{c_{m,n},q} = 0$, otherwise.

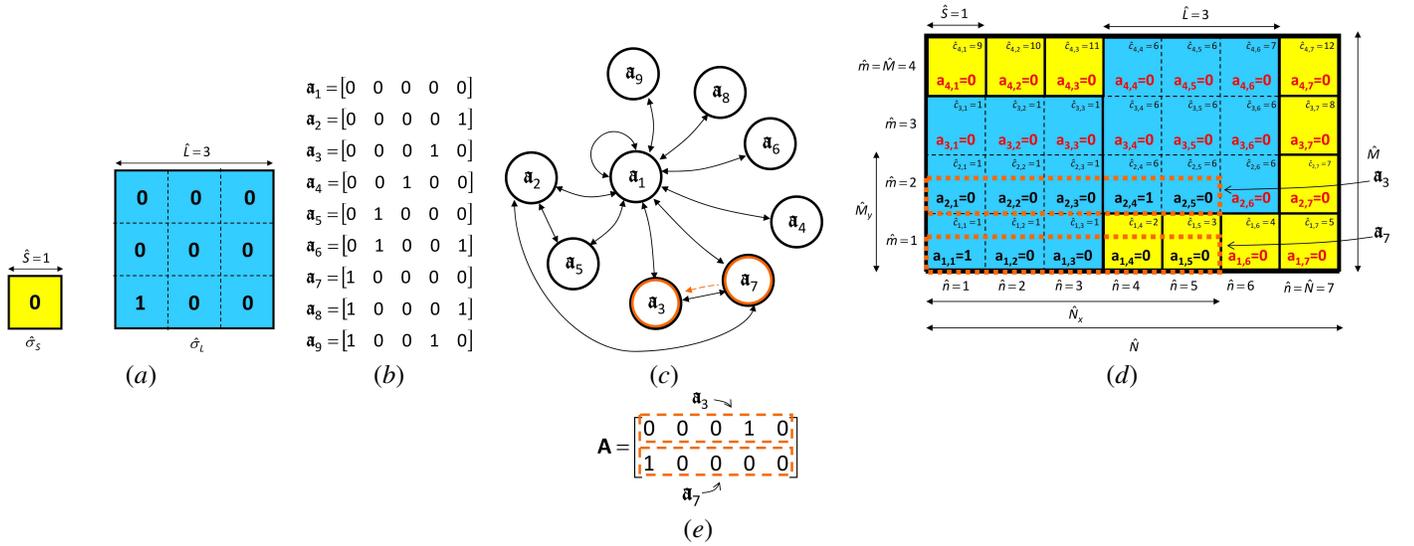

Figure 3.  *Solution Coding and Graph Mapping* ($M = 4$, $N = 7$; $S = 1$, $L = 3$) - (*a*) Coding of the $\sigma_S$ and $\sigma_L$ tiles [Fig. 1(*b*) - $R$ physical aperture] in the virtual aperture $\hat{R}$, (*b*) the row dictionary (i.e., the set of $H$ admissible configurations of a row of **A**), $\mathfrak{A} \triangleq [\mathbf{a}_h;$ $h = 1, ..., H]$, (*c*) the solution graph $\mathcal{G}$, (*d*) virtual representation of the tiling configuration (i.e., the tiling mapped in $\hat{R}$) with element-membership vector $\mathbf{c} = \{1, 1, 1, 2, 3, 4, 5, 1, 1, 1, 6, 6, 6, 6, 7, 1, 1, 6, 6, 8, 9, 10, 11, 6, 6, 6, 12\}$, and (*e*) its corresponding matrix **A** (i.e., a set of $\hat{M}_y$ elements of $\mathfrak{A}$).

The synthesis problem at hand then consists in defining the tiling configuration $\mathbf{c}$ and the set of phase coefficients, $\boldsymbol{\beta}$, so that the radiated power pattern $\mathbf{P}(\theta, \phi; \mathbf{c}, \boldsymbol{\beta})$ [$\mathbf{P}(\theta, \phi; \mathbf{c}, \boldsymbol{\beta}) \triangleq |\mathbf{E}(\theta, \phi; \mathbf{c}, \boldsymbol{\beta})|^2$] pointing along $(\theta_0, \phi_0)$

fits user-defined constraints on the sidelobe level (*SLL*) [$SLL \triangleq \frac{\max_{(\theta, \phi) \in \Xi} \{\mathbf{P}(\theta, \phi; \mathbf{c}, \boldsymbol{\beta})\}}{\mathbf{P}(\theta_0, \phi_0; \mathbf{c}, \boldsymbol{\beta})}$, $\Xi$ being the sidelobe region] and on the half-power beamwidth (*HPBW*) along the principal planes [i.e., the horizontal/azimuth ($\phi = 0$ [deg]) and the vertical/elevation ($\phi = 90$ [deg]) planes], $HPBW_{az}$ and $HPBW_{el}$: $SLL \leq \eta_{SLL}$, $HPBW_{az} \leq \eta_{az}$, and $HPBW_{el} \leq$

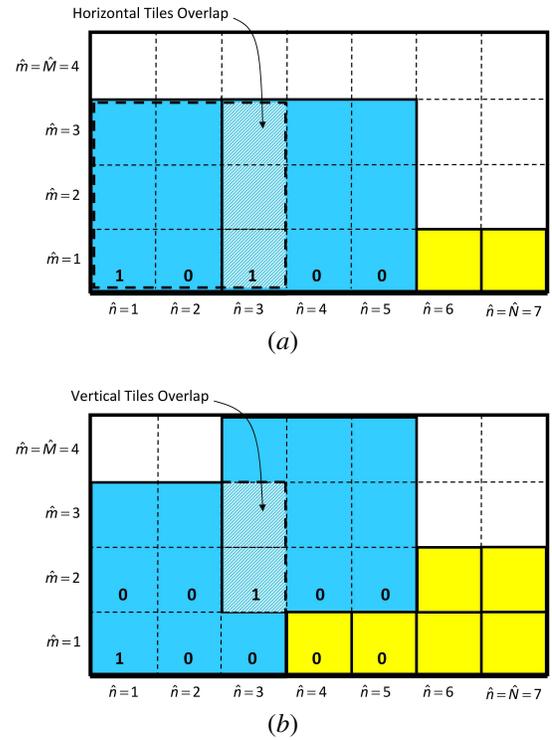

Figure 4.  *Solution Coding and Graph Mapping* ($M = 4$, $N = 5$; $S = 1$, $L = 3$) - Examples of (*a*) an horizontal overlap and (*b*) a vertical overlap between two $\sigma_L$ tiles in the virtual aperture $\hat{R}$.

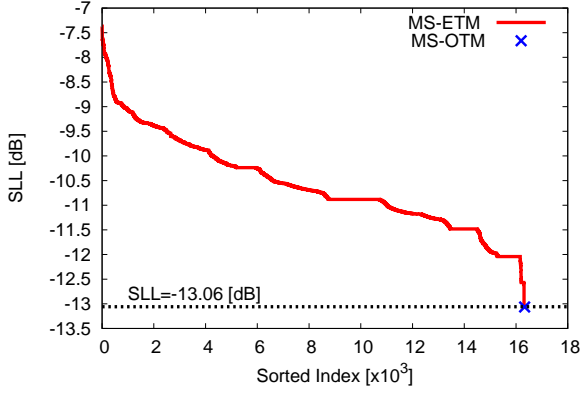

Figure 5.    *Numerical Assessment* ($M = 5$, $N = 8$, $d = 0.5\lambda$; $S = 1$, $L = 2$) - Value of the *SLL* of the whole set of $U_2$ tiling configurations sorted from the worst to the best ($Q = 34$: $Q_S = 32$ and $Q_L = 2$) and of the best *MS-OTM* solution ($P = 12$).

$\eta_{el}$.

Since pointing the main-lobe towards a desired steered direction ($\theta_0, \phi_0$) means analytically setting the values of the sub-array phases as follows

$$\beta_q = -\frac{2\pi}{\lambda}\left(x_q \sin\theta_0 \cos\phi_0 + y_q \sin\theta_0 \sin\phi_0\right),$$
$$q = 1, ..., Q \quad (3)$$

where $(x_q, y_q)$ are the planar coordinates of the center of the $q$-th ($q = 1, ...., Q$) tile given by

$$\begin{pmatrix} x_q \\ y_q \end{pmatrix} = \frac{1}{\gamma_q} \sum_{m=1}^{M} \sum_{n=1}^{N} \delta_{c_{mn}q} \begin{pmatrix} x_n \\ y_m \end{pmatrix} \quad (4)$$

where $\gamma_q = \gamma_S$ or $\gamma_q = \gamma_L$ if the $q$-th tile is either $\sigma_S$ or $\sigma_L$, respectively, then $\mathbf{P}\left(\theta, \phi; \mathbf{c}, \boldsymbol{\beta}\right)$ and $\mathbf{P}\left(\theta, \phi; \mathbf{c}\right)$ since $\boldsymbol{\beta} = \boldsymbol{\beta}\left(\mathbf{c}\right)$. The tiling problem can be then reformulated as follows:

*MSTA Synthesis Problem* - Given an aperture $R$, a rectangular phased array composed of $M$ rows, each containing $N$ elements, and two types of isophoric square tiles, $\sigma_S$ and $\sigma_L$, that contain $\gamma_S$ and $\gamma_L = \hat{L} \times \gamma_S$ ($\hat{L} \geq 2$) array elements, respectively, find the optimal $R$-complete tiling configuration, $\mathbf{c}^{opt}$ (i.e., the optimal arrangement of $\sigma_S$ and $\sigma_L$ sub-array modules) so that the pattern-fitting cost function $\Phi$

$$\Phi\left(\mathbf{c}\right) \triangleq w_{sl} \{|SLL\left[\mathbf{P}\left(\theta, \phi; \mathbf{c}\right)\right] - \eta_{SLL}|$$
$$\times \mathcal{H}\left[|SLL\left[\mathbf{P}\left(\theta, \phi; \mathbf{c}\right)\right] - \eta_{SLL}|\right]\}$$
$$+ w_{bw} \{|HPBW_{az}\left[\mathbf{P}\left(\theta, \phi; \mathbf{c}\right)\right] - \eta_{az}|$$
$$\times \mathcal{H}\left[|HPBW_{az}\left[\mathbf{P}\left(\theta, \phi; \mathbf{c}\right)\right] - \eta_{az}|\right]$$
$$+ |HPBW_{el}\left[\mathbf{P}\left(\theta, \phi; \mathbf{c}\right)\right] - \eta_{el}|$$
$$\times \mathcal{H}\left[|HPBW_{el}\left[\mathbf{P}\left(\theta, \phi; \mathbf{c}\right)\right] - \eta_{el}|\right]\} \quad (5)$$

is minimized (i.e., $\mathbf{c}^{opt} \triangleq \arg\{\min_{\mathbf{c}}\left(\Phi\left(\mathbf{c}\right)\right)\}$), $w_{sl}$ and $w_{bw}$ being real-valued user-defined weighting coefficients, while $\mathcal{H}\left[\cdot\right]$ is the Heaviside function.

## III. MULTI-SQUARE TILED ARRAY SYNTHESIS

The aforementioned *MSTA* synthesis problem is solved by means of an ad-hoc strategy that exploits the recent theory from [37] on the set of admissible complete partitionings of a rectangle by means of two squares of different sizes. More specifically, the proposed tiling method consists of the following logical steps: (*3.1*) Assessment of the condition for the complete tileability of the rectangular array aperture $R$ with the two tiles $\sigma_S$ and $\sigma_L$; (*3.2*) Binary coding of the $M$ rows composing the rectangular array aperture and mapping of the whole set of admissible tile-arrangements in a graph; (*3.3*) Searching for the optimal tiling, $\mathbf{c}^{opt}$, within the solution space by means of a computationally-effective optimization algorithm called *Multi-Square Optimization-Based Tiling Method* (*MS-OTM*). Each step will be detailed in the following.

### A. Aperture Covering Theorems

Let us model each $(m, n)$-th element of the array with a square tile so that the whole aperture $R$ is composed of $M \times N$ pixels. The array aperture $R$ can be fully covered by $\sigma_S$ and $\sigma_L$ tiles [Fig. 2(*a*)] if and only if the conditions

- $S$ divides both $M$ and $N$, namely $(M \bmod S) = 0$ and $(N \bmod S) = 0$, $mod$ being the modulo operation;
- $M$ (or $N$) is equal or larger than $L$ [$M \geq L$ (or $N \geq L$)] and $N$ (or $M$) is a positive linear combination of $S$ and $L$ [$N = w_1 S + w_2 L$ (or $M = w_1 S + w_2 L$), $w_1$ and $w_2$ being real-valued coefficients];

jointly hold true. Moreover, if none of the above conditions is verified, the region $R$ is not fully tileable with an arrangement of both $\sigma_S$ and $\sigma_L$ tiles [Fig. 2(*b*)]. Otherwise, if only the first condition holds true, then $R$ can be completely partitioned with only $\sigma_S$ tiles.

### B. Solution Coding and Graph-Mapping

Once the tileability of the aperture $R$ has been positively assessed, there is the need of a suitable coding for a trial tiling configuration as well as of a compact representation of the solution space of the admissible tiling solutions. Towards this end, the theory of [37] is here extended to the case of two square tiles fitting the condition $\gamma_L = \hat{L} \times \gamma_S$ ($\hat{L} \geq 2$). More in detail, a virtual aperture $\hat{R}$ of $\hat{M}$ [$\hat{M} = \frac{M}{S}$; $\hat{M} = 4$ - Fig. 3(*d*)] rows and $\hat{N}$ [$\hat{N} = \frac{N}{S}$; $\hat{N} = 7$ - Fig. 3(*d*)] columns, thus containing $\hat{M} \times \hat{N}$ pixels, is first defined by rescaling $S$ times the physical aperture $R$. Accordingly, the smallest tile $\sigma_S$ of dimensions $\gamma_S$ turns out being represented in the virtual aperture by a single-pixel ($\gamma_{\hat{S}} \triangleq \hat{S} \times \hat{S} = 1$ since $\hat{S} = \frac{S}{S}$) virtual tile $\hat{\sigma}_S$ [Fig. 3(*a*)], while the largest one, $\sigma_L$, is mapped into an equivalent virtual tile $\hat{\sigma}_L$ that contains $\gamma_{\hat{L}}$ ($\gamma_{\hat{L}} \triangleq \hat{L} \times \hat{L}$) pixels [$\hat{L} = 3$ - Fig. 3(*a*)].[1]

Given the virtual aperture, a trial tiling configuration is then encoded into a binary matrix $\mathbf{A} = \left\{a_{\hat{m},\hat{n}} \in [0, 1]; \hat{m} = 1, ..., \hat{M}_y; \hat{n} = 1, ..., \hat{N}_x\right\}$, being $\hat{M}_y \triangleq \hat{M} - \hat{L} + 1$ [$\hat{M}_y = 2$ - Fig. 3(*d*)] and $\hat{N}_x \triangleq \hat{N} - \hat{L} + 1$ [$\hat{N}_x = 5$ - Fig. 3(*d*)], where the value of the bits (i.e., the binary entries of the matrix) associated to each pixels/elements of $\hat{R}$ are set according to the following

---
[1] It is worth noticing that $R$ and $\hat{R}$ coincide when $S = 1$ and all coding/mapping operations can be directly performed on the physical aperture at hand.

Table I
NUMBER OF TILING CONFIGURATIONS, $U_2$.

| $M = N$ | $U_2$ |
|---|---|
| 4 | 35 |
| 8 | $12.72 \times 10^6$ |
| 12 | $60.71 \times 10^{15}$ |
| 16 | $36.41 \times 10^{29}$ |

*tile-placement rules* [Fig. 3(a)]: (*a*) the bit corresponding to the pixel of a virtual tile $\hat{\sigma}_S$ is set to zero (i.e., $a_{\hat{m},\hat{n}} = 0$); (*b*) the bit corresponding to the pixel in the bottom-left corner of a virtual tile $\hat{\sigma}_L$ is set to one (i.e., $a_{\hat{m},\hat{n}} = 1$), while the others, still belonging to the same tile $\hat{\sigma}_L$, are set to zero (i.e., $a_{\hat{m},\hat{n}} = 0$). It is worth highlighting that the binary matrix $\mathbf{A}$ has dimensions $\hat{M}_y \times \hat{N}_x$ instead of $\hat{M}$ ($M \geq \hat{M} > \hat{M}_y$) and $\hat{N}$ ($N \geq \hat{N} > \hat{N}_x$) since the top $\hat{L} - 1$ rows and the rightmost $\hat{L} - 1$ columns always contain zeros [Fig. 3(d)], as a consequence of the rules (*a*) and (*b*). Consequently, there is a significant advantage in preferring $\mathbf{A}$ to $\mathbf{c}$ for coding a trial tiling configuration because of the drastic reduction of the cardinality of the solution space from $U = Q^{M \times N}$ down to $U = 2^{\hat{M}_y \times \hat{N}_x}$. Successively, let us consider the $\hat{M}_y$ rows of the matrix $\mathbf{A}$ ($\mathbf{A} \triangleq \left[\mathbf{a}_{\hat{m}}; \hat{m} = 1, ..., \hat{M}_y\right]^T$, $\mathbf{a}_{\hat{m}} \triangleq \left\{a_{\hat{m},\hat{n}}; \hat{n} = 1, ..., \hat{N}_x\right\}$ being the $\hat{m}$-th ($\hat{m} = 1, ..., \hat{M}_y$) row of the matrix $\mathbf{A}$) [Fig. 3(e)] to define the space of all the admissible tilings [37]. Since each row contains $\hat{N}_x$ pixels/bits, one would assume that the row dictionary (i.e., the $H$ admissible configurations of a row of $\mathbf{A}$), $\mathfrak{A} \triangleq \{\mathbf{a}_h; h = 1, ..., H\}$, has a cardinality equal to $H = 2^{\hat{N}_x}$, including $\mathbf{a}_h = \mathbf{0}$ (i.e., the row containing all zeros, $\mathbf{0} = \left\{a_{\hat{m},\hat{n}} \triangleq 0; \hat{n} = 1, ..., \hat{N}_x\right\}$), but differently it is much smaller ($H < 2^{\hat{N}_x}$) since all strings that encode an overlap between two virtual tiles must be excluded [Figs. 4(a)-4(b)]. In order to avoid an horizontal overlap [Fig. 4(a)], the binary entries of the generic $h$-th ($h = 1, ..., H$) element of $\mathfrak{A}$ are required to always fit the *'existence conditions'* when $\mathbf{a}_{h,\hat{n}} = 1$ ($\hat{n} \in [1 : \hat{N}_x]$):

$$
\begin{cases}
\mathbf{a}_{h,i} = 0 \; for \\
\quad (\hat{n} + 1) \leq i \leq \left(\hat{n} + \hat{L} - 1\right) \; if \\
\qquad\qquad\qquad \hat{n} \in \left[1 : \left(\hat{L} - 1\right)\right] \\
\left(\hat{n} - \hat{L} + 1\right) \leq i \leq \left(\hat{n} + \hat{L} - 1\right) \; if \\
\qquad\qquad \hat{n} \in \left[\hat{L} : \left(\hat{N}_x - \hat{L} + 1\right)\right] \\
\left(\hat{n} - \hat{L} + 1\right) \leq i \leq \left(\hat{N}_x - \hat{n}\right) \; if \\
\qquad\qquad \hat{n} \in \left[\left(\hat{N}_x - \hat{L} + 2\right) : \hat{N}_x\right]
\end{cases}
. \quad (6)
$$

According to (6), the row dictionary $\mathfrak{A}$ [Fig. 3(b)] and the set of nodes of the rows-graph [Fig. 3(c)] are then defined. However, it is not possible to iterate $\hat{M}_y$-times the assignment of a randomly-chosen $h$-th ($h \in [1 : H]$) element of $\mathfrak{A}$, $\mathbf{a}_h$, to a generic $\hat{m}$-th ($\hat{m} = 1, ..., \hat{M}_y$) row $\mathbf{a}_{\hat{m}}$ to generate a trial matrix $\mathbf{A}$ (i.e., a trial array tiling) because of possible vertical overlaps between two $\hat{\sigma}_L$ tiles [e.g., Fig. 4(b)]. To avoid such an unphysical arrangement when filling the matrix $\mathbf{A}$ with the elements of the dictionary $\mathfrak{A}$, the following *'feasibility assignment condition'* is stated: "*If the $\hat{m}$-th row of $\mathbf{A}$ contains an one-bit at the $\hat{n}$-th pixel position* (i.e., $a_{\hat{m},\hat{n}} = 1$, $\hat{n} \in [1 : \hat{N}_x]$) *and*

$$
\mathbf{a}_{h,\hat{n}} = 0 \quad (7)
$$

*as well as* (6) *hold true, then any $h$-th ($h \in [1 : H]$) element of the dictionary, $\mathbf{a}_h$, can be assigned to any row of (a) the $\left(\hat{L} - 1\right)$ superior ones when $\hat{m} \in \left[1 : \left(\hat{M}_y - \hat{L} + 1\right)\right]$ and of (b) the $\left(\hat{M}_y - \hat{m}\right)$ ones, from the $(\hat{m}+1)$-th row until the (last) one (i.e., the $\hat{M}_y$ row), when $\hat{m} \in \left[\left(\hat{M}_y - \hat{L} + 2\right) : \hat{M}_y\right]$ (i.e., one of the top $\hat{L} - 1$ rows of $\mathbf{A}$).*"

Thanks to this formalism and as in [37], the solution space of the admissible tilings can be then represented through a solution graph $\mathcal{G}$ [Fig. 3(c)] where (*i*) each node corresponds to an element of the row dictionary $\mathfrak{A}$ that satisfies the *existence conditions* and (*ii*) two nodes of $\mathcal{G}$ are connected only if the corresponding rows fit the *feasibility assignment condition*. Therefore, a matrix $\mathbf{A}$ turns out to be equivalent to a path of $\hat{M}_y$ nodes in $\mathcal{G}$ [Fig. 3(c)] with a node visited more than one time, as well.

Once the matrix $\mathbf{A}$ (i.e., a path of connected nodes in the graph $\mathcal{G}$) has been determined, the corresponding arrangement of $\hat{\sigma}_S$ and $\hat{\sigma}_L$ virtual tiles is set in $\hat{R}$ according to the *tile-placement rules* [Fig. 3(b)] so that a bit equal to one in the $(\hat{m},\hat{n})$-th entry of the matrix $\mathbf{A}$ [e.g., $a_{2,4} = 1$ - Fig. 3(e)] indicates the presence of a tile of larger size $\hat{\sigma}_L$ with bottom-left corner at the $(\hat{m},\hat{n})$-th position of the virtual array $\hat{R}$ [e.g., $a_{2,4} = 1$ - Fig. 3(d)]. Finally, the tiling configuration in the physical aperture $R$ is yielded by "stretching" $S$ times the sizes of the virtual tiles of $\hat{R}$ along $x$ and $y$ directions. Mathematically, it means setting the membership of the $(p,q)$-th [$p \triangleq \hat{m} + (i - 1) \times S$ and $q \triangleq \hat{n} + (j - 1) \times S$; $i = 1, ..., S$; $j = 1, ..., S$] elementary radiator of the array to the same value of the membership of the $(\hat{m},\hat{n})$-th ($\hat{m} = 1, ..., \hat{M}_y$; $\hat{n} = 1, ..., \hat{N}_x$) pixel of the virtual aperture:

$$
c_{p,q} = \hat{c}_{\hat{m},\hat{n}}. \quad (8)
$$

Concerning the number of tiling configurations of the array aperture, the dimension of the solution space $U$ has been derived in closed form [37] for the case $\hat{L} = 2$:

$$
U_2 = \mathbf{1}^T \mathbf{G}^{\hat{N}-2} \mathbf{1} \quad (9)
$$

where $\mathbf{1}$ is the identity vector of length $H$, $T$ stands for the transpose operator, and $\mathbf{G}$ is the adjacent matrix of the graph $\mathcal{G}$ [38] (see *Appendix I*). For instance, let us consider the values of $U_2$ for square apertures up to $\hat{M} = \hat{N} = 16$ reported in Tab. I. As it can be noticed, the cardinality of the space of clustered layouts grows exponentially, thus the use of enumerative/brute-force algorithms testing all the possible arrangements is possible only for very small apertures. Therefore, an optimization-based synthesis strategy is used to deal with large aperture, as well.

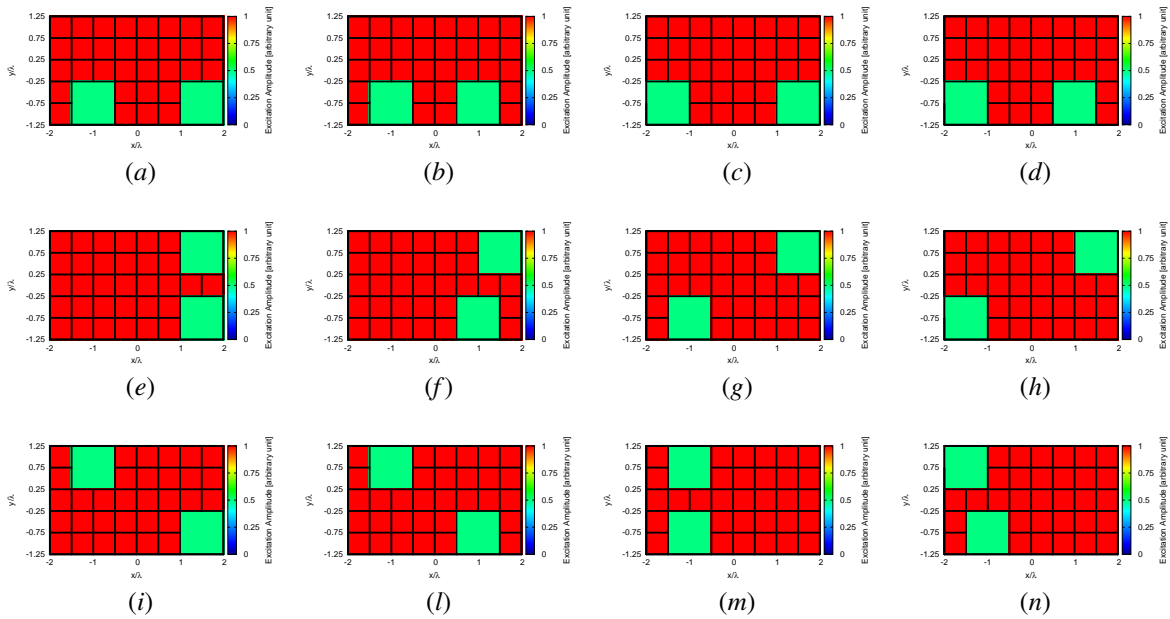

Figure 6.   *Numerical Assessment* ($M = 5$, $N = 8$, $d = 0.5\lambda$; $S = 1$, $L = 2$; $Q = 34$) - Two-level distributions of the equivalent element-level amplitude excitations $\{\alpha_{m,n}^{(q)}; m = 1, ..., M; n = 1, ..., N; q = 1, ..., Q\}$, of the global best tiling configurations ($Q_S = 32$ and $Q_L = 2$).

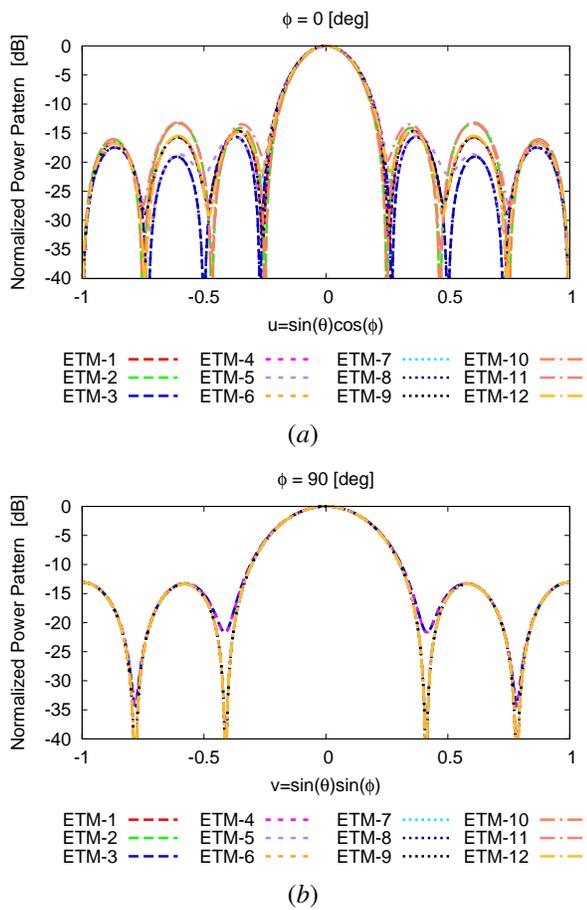

Figure 7.   *Numerical Assessment* ($M = 5$, $N = 8$, $d = 0.5\lambda$; $S = 1$, $L = 2$; $Q = 34$: $Q_S = 32$, $Q_L = 2$) - Cuts of the normalized power pattern along the principal planes, (*a*) $\phi = 0$ [deg] and (*b*) $\phi = 90$ of the global best tiling configurations in Fig. 6.

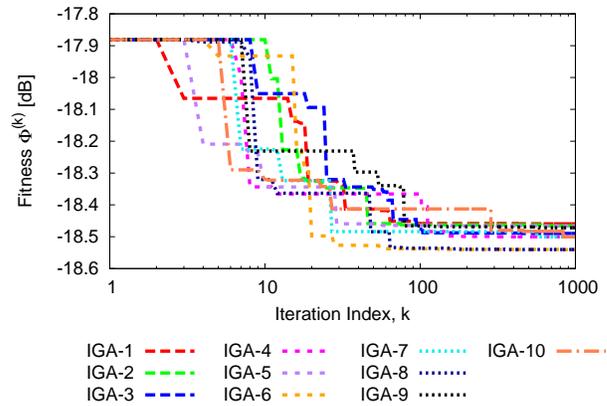

Figure 8.   *Numerical Assessment* ($M = 15$, $N = 20$, $d = 0.5\lambda$; $S = 1$, $L = 2$; $P = 42$, $\rho_c = 0.9$, $\rho_m = 0.01$, $K = 10^3$; $O = 10$) - Behaviour of the fitness of the best individual, $\Phi_k^{opt}$, versus the iteration index $k$ ($k = 1, ..., K$) for each $o$-th ($o = 1, ..., O$) *MS-OTM* (*IGA*-based) run.

### C. Multi-Square Optimization-Based Tiling Method (MS-OTM)

The goal of using an heuristic optimization approach for minimizing (5) is dual. First, the reduction of the number of functional evaluation with respect to an exhaustive search to enable the synthesis of large apertures, which are generally considered in realistic applications. Second, the convergence - with high probability - to the optimal tiling (i.e., $\mathbf{c}^{opt} = \arg \{\min_{\mathbf{c}_u} [\Phi(\mathbf{c}_u); u = 1, ..., U]\}$) or to a solution fitting the requirements within a user-defined tolerance $\eta$ [i.e., $\mathbf{c}^{opt}$: $\Phi(\mathbf{c}^{opt}) \leq \eta$]. Towards this end, the proposed *MS-OTM* is based on an integer-coded Genetic Algorithm (*IGA*) where a trial tiling is coded into an integer string $\mathbf{t} = \left\{ t_{\hat{m}} \in [1 : H]; \hat{m} = 1, ..., \hat{M}_y \right\}$ whose $\hat{m}$-th entry is the index of an element of the row dictionary $\underline{\mathfrak{A}}$ [Fig. 3(b)] as well as a node of the solution graph $\mathcal{G}$ [Fig. 3(c)]. More specifically, the following procedural steps are performed:

- **Step 0:** *Virtual Aperture Set-up* - If $S > 1$, then rescale $S$ times the size of the aperture $R$ to yield the virtual aperture $\hat{R}$ (i.e., $M \to \hat{M}$ and $N \to \hat{N}$) as well as the area of the two-sizes tiles (i.e., $\sigma_S \to \hat{\sigma}_S$ and $\sigma_L \to \hat{\sigma}_L$);

- **Step 1:** *Solution Space Generation* - Generate the row dictionary $\underline{\mathfrak{A}}$ with $H$ elements of length $\hat{N}_x$ that satisfy the *existence condition*s (6). Execute the *IGA*-based iterative ($k$ being the iteration index) optimization loop:

- **Step 2:** *Population Initialization* ($k = 0$) - Randomly generate $P$ individuals, $\{ \mathbf{t}_k^{(p)} \}_{k=0}$; $p = 1, ..., P\}$, as follows

$$\left\{ t_{\hat{m},k}^{(p)} \right\}_{k=0} = r\left(\hat{m}\right); \ \hat{m} = 1, ..., \hat{M}_y \right\}; \qquad (10)$$
$$p = 1, ..., P$$

$r\left(\hat{m}\right)$ being an integer number randomly-selected within the interval $[1 : H]$ with uniform probability. If the rows of the $p$-th ($p \in [1 : P]$) individual do not satisfy the *feasibility assignment condition* (i.e., the $\hat{M}_y$ nodes coded in $\mathbf{t}_k^{(p)}$ are not connected by an edge in the solution graph $\mathcal{G}$), then replace it with another one - still randomly-generated - until the initial population is completed with all $P$ feasible tiling configurations;

- **Step 3:** *Fitness Evaluation* - Define the tiling matrix $\mathbf{A}_k^{(p)}$ ($p = 1, ..., P$) from the individual $\mathbf{t}_k^{(p)}$ ($p = 1, ..., P$) by setting each $\hat{m}$-th ($\hat{m} = 1, ..., \hat{M}_y$) row as follows

$$\mathbf{a}_{\hat{m}} \big]_k = \mathbf{a}_{t_{\hat{m},k}^{(p)}} . \qquad (11)$$

Generate the corresponding trial clustering $\mathbf{c}_k^{(p)}$ ($p = 1, ..., P$) on the physical aperture $R$ by first positioning the $\hat{\sigma}_S$ and $\hat{\sigma}_L$ virtual tiles in the virtual aperture $\hat{R}$ according to the *tile-placement rules* and then mapping the arising membership vector $\hat{\mathbf{c}}_k^{(p)}$ on $R$ by applying (8). Afterwards, set the fitness of each $p$-th ($p = 1, ..., P$) trial solution, $\Phi_k^{(p)}$, to the value of (5)

$$\Phi_k^{(p)} = \Phi\left(\mathbf{c}_k^{(p)}\right) \qquad (12)$$

$\beta_k^{(p)}$ being computed through (3). If $k = 0$, then go to Step 5;

- **Step 4:** *Reproduction Cycle* - Apply the *roulette-wheel* selection, the *single-point* crossover with probability $\rho_c$, and the mutation with probability $\rho_m$ [39] - properly customized to deal with integer-coded chromosomes - to generate a new population, $\{\mathbf{t}_k^{(p)}; p = 1, ..., P\}$. For each $p$-th integer string $\mathbf{t}_k^{(p)}$ check whether the *feasibility assignment condition* holds true, otherwise update the new trial solution by re-applying the *IGA* operators. Enforce the elitism mechanism [39] to keep the best solution found so far within the current $k$-th population ($\mathbf{c}_k^{best} = \arg\left\{\min_{\mathbf{c}_i^{(p)}}\left[\Phi\left(\mathbf{c}_i^{(p)}\right); p = 1, ..., P\right]; l = 1, ..., k\right\}$);

- **Step 5:** *Convergence Check* - Stop the iterative process if the maximum number of iterations $K$ is reached (i.e., $k = K$) or the fitness of the best individual $\Phi_k^{opt}$ ($\Phi_k^{opt} \triangleq$ $\min_{p=1,...,P}\left\{\Phi_k^{(p)}\right\}$) is below $\eta$, then output $\mathbf{c}^{opt}$ ($\mathbf{c}^{opt} = \mathbf{c}_k^{best}$) and $\boldsymbol{\beta}^{opt}$ (3). Otherwise, update the iteration index ($k \leftarrow k + 1$) and go to Step 3.

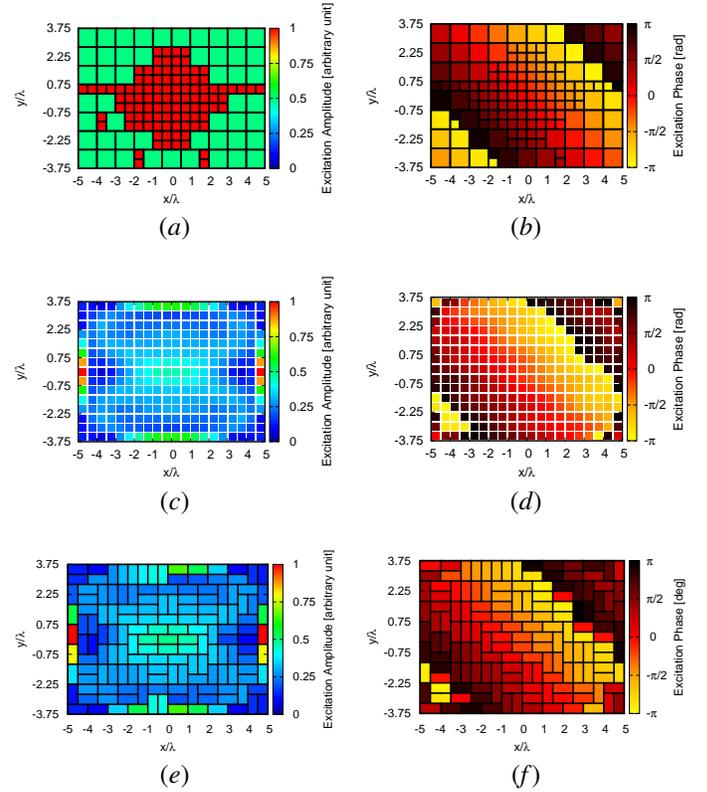

Figure 9. *Numerical Assessment* ($M = 15$, $N = 20$, $d = 0.5\lambda$, $(\theta_0, \phi_0) = (8, 45)$ [deg]; $S = 1$, $L = 2$) - Distribution of the element-level $(a)(c)(e)$ amplitude and $(b)(d)(f)$ phase excitations of $(a)(b)$ the *MS-OTM* ($P = 42$, $\rho_c = 0.9$, $\rho_m = 0.01$, $K = 10^3$) optimized *MSTA* ($Q = 150$: $Q_L = 50$, $Q_S = 100$), $(c)(d)$ the reference *FPA* ($SLL = -20$ [dB]), and $(e)(f)$ the *D-OTM* optimized ($P = 326$, $\rho_c = 0.9$, $\rho_m = 0.01$, $K = 10^3$) *DTA* ($Q = 150$: $Q_H = 82$, $Q_V = 68$).

## IV. NUMERICAL VALIDATION

The proposed isophoric multi-square tiled architecture and the *MS-OTM* synthesis strategy are analyzed and assessed in a representative set of numerical examples concerned with different apertures and design constraints. More specifically, the *MS-OTM* approach is firstly validated in a small array benchmark case to prove its effectiveness and reliability to converge towards the global-optimum/best-solution of the optimization/synthesis problem at hand. Then, a comparison with different tiling layouts composed of domino tiles with non-isophoric feeding [29] is presented to give some indications on the pros and the cons of adopting an isophoric multi-square tiling of the array aperture. Afterwards, two realistic antenna designs for space applications are reported and discussed.

The first example deals with a rectangular aperture hosting $M \times N = 5 \times 8$ isotropic [i.e., $\mathbf{e}_{m,n}(\theta, \phi) = 1$] elements spaced by $d_x = d_y = 0.5\lambda$, while the two square tiles, $\sigma_s$ and $\sigma_l$, contain $\gamma_S = 1 \times 1$ and $\gamma_L = 2 \times 2$ elements (i.e., $S = 1$ and $L = 2$) so that the physical aperture $R$ coincides with the virtual one $\hat{R}$. As for the goal of the synthesis, it has been chosen to minimize the *SLL* of the radiated power pattern when pointing along broadside. Accordingly, the weighting coefficients and the *SLL* constraint in (5) have been set to $w_{sl} = 1$, $w_{bw} = 0$, and $\eta_{SLL} = 0$, respectively.

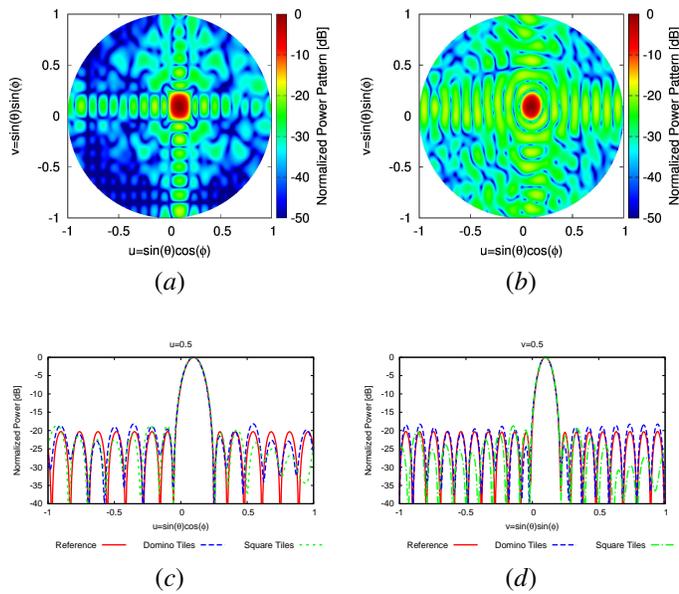

Figure 10. *Numerical Assessment* ($M = 15$, $N = 20$, $d = 0.5\lambda$, $(\theta_0, \phi_0) = (8, 45)$ [deg]) - Color-maps of the normalized power pattern radiated by (*a*) the *MS-OTM* optimized ($P = 42$, $\rho_c = 0.9$, $\rho_m = 0.01$, $K = 10^3$) *MSTA* ($S = 1$, $L = 2$; $Q = 150$: $Q_L = 50$, $Q_S = 100$) and (*b*) the *D-OTM* optimized ($P = 326$, $\rho_c = 0.9$, $\rho_m = 0.01$, $K = 10^3$) *DTA* ($Q = 150$: $Q_H = 82$, $Q_V = 68$). Cuts of the normalized power patterns of the *MSTA*/*DTA*/*FPA* along (*c*) the $u = 0.5$ and (*d*) the $v = 0.5$ planes.

Since the number of elements of the row dictionary $\mathfrak{A}$ is equal to $H = 34$, the cardinality of the solution space turns out to be computationally tractable [$U_2 = 16334$ (9)]. Thus, the whole set of $U_2$ admissible aperture-tilings has been generated and an exhaustive enumerative evaluation has been performed to identify the global optimum tiling. More specifically, the Multi-Square Exhaustive Tiling Method (*MS-ETM*) [40] has been used to exhaustively explore the solution graph $\mathcal{G}$ by computing the fitness of all array partitionings, as well. On an Intel 1.6GHz i5CPU, 8GB of RAM computer platform, the generation of all admissible tilings required less than 2 seconds, while the evaluation of the corresponding cost function values took 1 : 43 [h:min] by sampling the angular $(u, v)$ domain with $512 \times 512$ samples ($u \triangleq \sin\theta\cos\phi$, $u \triangleq \sin\theta\sin\phi$). Figure 5 shows the fitness of the array layouts sorted from the worst to the best. This latter corresponds to a *SLL* equal to $SLL^{opt} = -13.06$ [dB] and it has been reached by 24 different tiling configurations, all having $Q = 34$ tiles: $Q_S = 32$ of type $\sigma_S$ and $Q_L = 2$ of type $\sigma_L$. The most representative ones (i.e., the remaining best arrangements can be obtained from these by either a clockwise or a counter-clockwise rotation of 180 [deg]) are reported in Fig. 6 by plotting the two-level distribution of the equivalent element-level amplitude excitations $\{\alpha_{m,n}^{(q)}; \ m = 1, ..., M; \ n = 1, ..., N; \ q = 1, ..., Q\}$. For completeness, the cuts of the power pattern along the principal planes, $\phi = 0$ [deg] [Fig. 7(*a*)] and $\phi = 90$ [deg] [Fig. 7(*b*)], are given in Fig. 7. The proposed *MS-OTM* approach has been then run to evaluate the successful rate of the *IGA* to converge to one among the best solutions. More in detail, three different population sizes with $P = \{1; 2; 3\} \times V$ individuals, $V$ ($V \triangleq \hat{M}_y$, $\hat{M}_y = 4$) being the number of problem unknowns, have been taken into account and the other *IGA* parameters have been set to $K = 100$ (maximum number of iterations), $\rho_c = 0.9$ (crossover probability), $\rho_m = 0.01$ (mutation probability) [39]. Due to the stochastic nature of *GA*s, $O = 100$ different optimizations have been executed for each setup of the control parameters and the results are summarized in Tab. II. As expected, the success rate increases with the population size and the optimization process always converges to low values of the cost function (worst/max $SLL = -12.04$ [dB] - Tab. II) within the whole set of admissible values (Fig. 5). When using a population of $P = 12$ (i.e., $P = 3 \times V$), the *MS-OTM* is always able to find an optimal solution (Tab. II) with $P \times K = 1200$ evaluations of the cost function. Such an amount corresponds to the 7.3% of the cardinality of the solution space, $U$, and it points out the non-negligible saving of *CPU*-time with respect to the *MS-ETM* without loss of optimization performance.

The second example is devoted to compare the *MSTA* architecture, synthesized with the *MS-OTM* approach, with the *Domino-Tiled Array* (*DTA*) layout designed with the optimization-based tiling method (*D-OTM*) presented in [29] where single-shaped domino-like tiles were used. Towards this end, a rectangular array of $M \times N = 15 \times 20$ isotropic elements spaced by $d_x = d_y = 0.5 \lambda$ has been taken into account. As in the previous test case, square tiles with $S = 1$ and $L = 2$ as well as the minimization of the *SLL* (i.e., $\Phi(\mathbf{c}) \triangleq SLL[\mathbf{P}(\theta, \phi; \mathbf{c})] \times \mathcal{H}[|SLL[\mathbf{P}(\theta, \phi; \mathbf{c})]|]$), but along the pointing direction $(\theta_0, \phi_0) = (8, 45)$ [deg], have been chosen to complete the definition of the benchmark. Moreover, the maximum number of sub-arrays has been limited to $\frac{M \times N}{2}$ (i.e., $Q \leq \frac{M \times N}{2}$) for a fair comparison between the *MSTA* arrangement and the *DTA* one [29].

Concerning the *MSTA* design, since the number of nodes in the solutions graph $\mathcal{G}$ turns out to be equal to $H = 10964$, the solution space has a dimension, $U_2 > 10^{30}$, which is out of the computational feasibility of the *EM-ETM*, thus the synthesis process has been performed only with the *MS-OTM*. Accordingly, a set of $O = 10$ different *IGA* runs have been executed by setting the *IGA* parameters as in the first example (i.e., $P = 3 \times V$, $\rho_c = 0.9$, and $\rho_m = 0.01$) except for the number of iterations ($K = 10^3$ vs. $10^2$), because of the wider dimension of the aperture at hand and the corresponding higher dimensionality of the solution space.

Table II
*Numerical Assessment* ($M = 5$, $N = 8$, $d = 0.5\lambda$; $S = 1$, $L = 2$) - SLL Statistics.

|  | Min [dB] | Max. [dB] | Avg. [dB] | Var. - | Success Rate [%] |
|---|---|---|---|---|---|
| *MS-ETM* | $-13.06$ | $-7.37$ | $-10.57$ | $0.85$ | $-$ |
| *MS-OTM* - $P = 4$ | $-13.06$ | $-12.04$ | $-12.51$ | $0.24$ | $42$ |
| *MS-OTM* - $P = 8$ | $-13.06$ | $-12.04$ | $-12.83$ | $0.18$ | $78$ |
| *MS-OTM* - $P = 12$ | $-13.06$ | $-13.06$ | $-13.06$ | $0.00$ | $100$ |

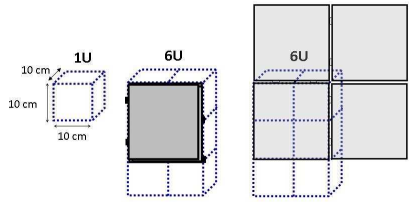

Figure 11. Sketches of a deployable implementation of a $4 \times 4$ *MSTA* on a 6-Units Cube-Sat payload [5].

Table III
*Numerical Assessment* ($M = 15$, $N = 20$, $d = 0.5\lambda$, $(\theta_0, \phi_0) = (8, 45)$ [DEG]; $S = 1$, $L = 2$) - PATTERN AND LAYOUT FEATURES.

| | $SLL$ [dB] | $D$ [dBi] | $HPBW_{az}$ [deg] | $HPBW_{el}$ [deg] | $TRM$ - |
|---|---|---|---|---|---|
| *FPA* | $-20.00$ | $27.69$ | $5.44$ | $7.34$ | $300$ |
| *MSTA* | $-18.54$ | $28.76$ | $5.60$ | $7.50$ | $150$ |
| *DTA* | $-18.12$ | $27.11$ | $5.49$ | $7.35$ | $150$ |

Figure 8 shows the evolution of the fitness of the best individual, $\Phi_k^{opt}$ ($k = 1, ..., K$) for each $o$-th ($o = 1, ..., O$) simulation. It can be noticed that the values of the *SLL* at the convergence ($k = K$) are very close to that of the best run (i.e. *GA*-8, $\Phi^{opt} = -18.54$ [dB]) when the *MSTA* is composed of $Q_L = 50$ tiles of type $\sigma_L$ and $Q_S = 100$ tile of type $\sigma_S$ ($Q = 150$), while the equivalent element-level amplitudes, $\{\alpha_{m,n}^{(q)}; m = 1, ..., M; n = 1, ..., N; q = 1, ..., Q\}$, and phases, $\{\beta_{m,n}^{(q)} = \beta_q$ (3); $m = 1, ..., M; n = 1, ..., N; q = 1, ..., Q\}$, are distributed as in Fig. 9($a$) and Fig. 9($b$), respectively, to afford the power pattern shown in Fig. 10($a$). As expected, the larger tiles are located at the edges of the aperture, while the smaller clusters turn out to be at the center to yield a tapering of the amplitude distribution over the antenna aperture so that $SLL = -18.54$ [dB] and the directivity is equal to $D = 28.76$ [dBi] (Tab. III).

As for the *DTA*, the binary *GA*-based *D-OTM* [29] has been applied with the same parameter setup of the *MS-OTM*, but using a larger population (i.e., $P = 326$) because of the greater length of the binary-encoded chromosomes since here the number of unknowns amounts to $V_{DTA} = 747$ (vs. $V_{MSTA} = 14$). Moreover, the non-isophoric sub-array feeding of the *DTA* has been determined through [41] by approximating the excitations [Figs. 9($c$)-9($d$)] of a reference *FPA* of $M \times N$ elements affording a beam steered towards $(\theta_0, \phi_0) = (8, 45)$ [deg] with $SLL = -20$ [dB]. The resulting domino-tiles arrangement and the corresponding amplitude and phase distributions are shown in Figs. 9($e$)-($f$), while the radiated power pattern is plotted in Fig. 10($b$). By comparing the pattern descriptors in Tab. III, it is worth noting that the domino-tiled array affords higher secondary lobes ($SLL^{DTA} = -18.12$ [dB] vs. $SLL^{MSTA} = -18.54$ [dB] - Tab. III) as well as a lower directivity ($D^{DTA} = 27.11$ [dBi] vs. $D^{MSTA} = 28.76$ [dBi] - Tab III). Indeed, the use of clusters of radiating elements and the beam broadening cause a decrease of directivity when focusing the beam out of the broadside direction which in this case amounts to $D_{loss} = 0.21$ [dBi], being $D^{MSTA} = 28.97$ [dBi] the directivity of the broadside directed beam when $(\theta_0, \phi_0) = (0, 0)$ [deg]. For the sake of completeness, the cuts of the *DTA*/*MSTA* power patterns along the principal planes crossing the main lobe are reported in Fig. 10($c$) [$u = 0.5$] and Fig. 10($d$) [$v = 0.5$].

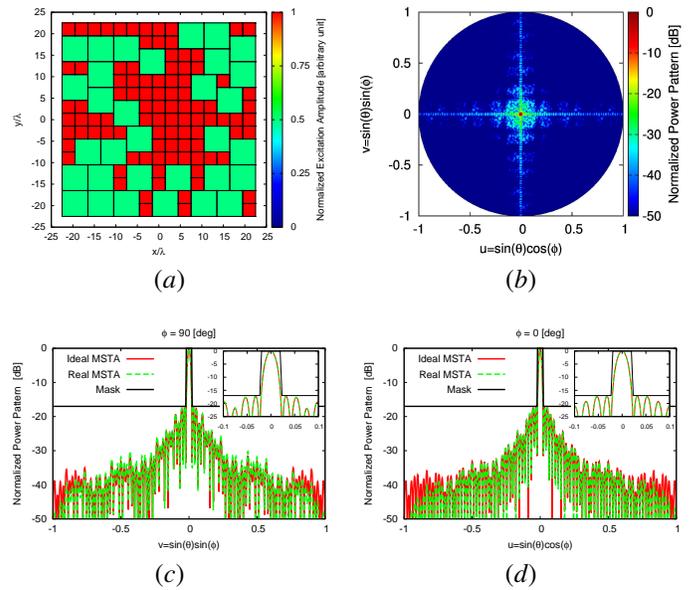

Figure 12. *Numerical Assessment* ($M = 90$, $N = 90$, $d = 0.5\lambda$, $(\theta_0, \phi_0) = (0.0, 0.0)$ [deg]; $S = 6$, $L = 12$; $Q = 129$: $Q_L = 32$, $Q_S = 97$) - Plots of ($a$) the equivalent element-level amplitude distribution and of the normalized power pattern radiated ($b$) in the whole angular range ($-1 \leq u \leq 1$, $-1 \leq v \leq 1$) and along ($c$) the $\phi = 90$ [deg] and ($d$) the $\phi = 0$ [deg] planes.

Table IV
*Numerical Assessment* ($M = 90$, $N = 90$, $d = 0.5\lambda$, $(\theta_0, \phi_0) = (0.0, 0.0)$ [DEG]) - PATTERN FEATURES.

| | $SLL$ [dB] | $HPBW_{az}$ [deg] | $HPBW_{el}$ [deg] | $G$ [dBi] |
|---|---|---|---|---|
| *Requirements* | $< -17.00$ | $< 1.20$ | $< 1.20$ | $> 42.00$ |
| *Ideal MSTA* | $-17.17$ | $1.17$ | $1.17$ | $43.51$ |
| *Real MSTA* | $-17.04$ | $1.16$ | $1.17$ | $43.60$ |

The synthesis problem addressed in the third example has been stated starting from the design guidelines for a precipitation radar system used in a recent CubeSat scientific mission [5]. More specifically, the requirements @ 35.75 [GHz] were: ($a$) $G > 42$ [dBi], $G$ being the broadside beam peak gain; ($b$) $HPBW < 1.2$ [deg] along both azimuth and elevation; (c) $SLL < -17$ [dB] (Tab. IV). Accordingly, a square aperture $R$ of $M \times N = 90 \times 90$ $\frac{\lambda}{2}$-spaced elements has been considered (i.e., $R$ extends over an area of $44.5\lambda \times 44.5\lambda$) for a twofold reason. On the one hand, to assure an ideal directivity of $D \simeq 44.0$ [dBi] when uniformly-tapering the excitations of an array of isotropic sources. On the other, to keep the whole array size (i.e., $4 \times 4$ Units) suitable for a deployable implementation on a 6-Unit cube-sat payload [5] as shown in Fig. 11. Moreover, the number of elementary radiators of the two *MSTA* tiles has been chosen equal to $\gamma_S = 6 \times 6$ and $\gamma_L = 12 \times 12$ (i.e., $S = 6$, $L = 12 \rightarrow \hat{L} = 2$; $V \triangleq \hat{M}_y = 14$), respectively.

Just running the *MS-OTM* $O = 10$ times, a solution with cost function value equal to $\Phi^{opt} = 4.3 \times 10^{-2}$ has been found whose radiated power pattern [Fig. 12($b$)] has $SLL = -17.17$ [dB] and $HPBW_{az} = HPBW_{el} = 1.17$ [deg], while the directivity turns out to be $D = 43.51$ [dB] (Tab. IV). The optimized tiling configuration contains $Q = 129$ tiles ($Q_L = 32$ and $Q_S = 97$) as shown in Fig. 12(*a*) where the equivalent element-level amplitudes are reported. For completeness, the top-view representation of the power pattern and its behavior along the two principal cuts are shown in Fig. 12(*b*) and Figs. 12(*c*)-(*d*), respectively.

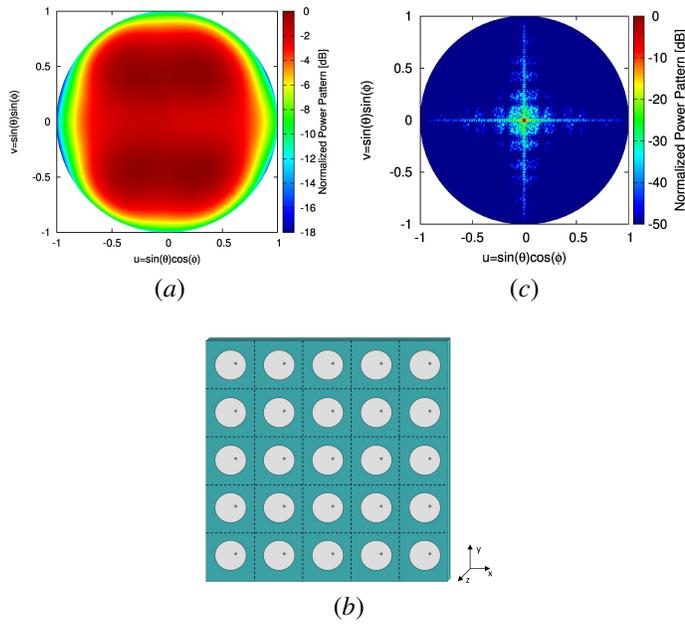

Figure 13. *Numerical Assessment (Real MSTA; M = 90, N = 90, d = 0.5λ, (θ₀,φ₀) = (0.0, 0.0) [deg]; S = 6, L = 12; Q = 129: Q_L = 32, Q_S = 97)* - Plots of (*a*) the embedded element pattern of the real elementary-radiator in the presence of (*b*) 5 × 5 equal neighbouring array-elements and of (*c*) the normalized power pattern radiated by the *MSTA*.

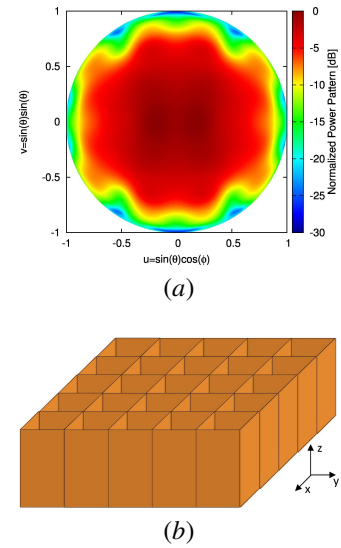

Figure 14. *Numerical Assessment (Real MSTA; d = 0.84λ)* - Plot of (*a*) the embedded element pattern of the real elementary-radiator in the presence of (*b*) 5 × 5 equal neighbouring square open-ended waveguides.

The *MSTA* layout in Fig. 12(*a*) has been also validated in a realistic implementation when non-ideal radiating elements are used. The antenna has been assumed to be an array of linearly polarized pin-fed circular patches working in the Ka-band around 35.75 [GHz] with patch diameter equal to 2.73 [mm], feed offset set to 0.31 [mm], and a slab of thickness 0.25 [mm] having $\tan\delta = 1.5 \times 10^{-3}$ and a relative permittivity $\epsilon = 2.94$. The mutual coupling effects as well as other electromagnetic phenomena among the array elements have been modeled by computing the embedded element pattern $\mathbf{e}(\theta,\phi)$ [Fig. 13(*a*)], which has been set identical for all radiators [i.e., $\mathbf{e}_{mn}(\theta,\phi) = \mathbf{e}(\theta,\phi)$; $m = 1,...,M$; $n = 1,...,N$], in the presence of two rings of neighboring elements [Fig. 13(*b*)]. The full-wave simulated radiation pattern is given in Fig. 13(*c*), while the values of the power pattern descriptors are reported in Tab. IV for a quick comparison with those from the ideal array and the project requirements. The outcomes are very positive since the application requirements are still satisfied (Tab. IV) and the pattern deviations from the ideal case are minimal [Fig. 13(*c*) vs. Fig. 12(*b*); Figs. 12(*c*)-12(*d*)] especially around the main beam zone as pointed out in the insets of Fig. 12(*c*) [$-0.1 \leq v \leq 0.1$ ($\phi = 90$ [deg])] and Fig. 12(*d*) [$-0.1 \leq u \leq 0.1$ ($\phi = 0$ [deg])].

The last test case deals with the synthesis of an isophoric phased array for satellite communications with the same requirements of [14][22], but exploiting a square aperture instead of a circular one. More in detail, the maximum antenna size, namely its extension along the diagonal, has been constrained to be smaller than 120 λ @ 19.0 [GHz]. Moreover, the array has been designed to scan the beam within the range $0 \leq \theta_0 \leq 1.12$ [deg] with a directivity pattern fitting the following conditions: (*i*) $D(\theta,\phi) \geq 43.8$ [dBi] for $0 \leq \theta \leq \theta^{EoC}$, $\theta^{EoC}$ being the angle at the edge-of-coverage (*EoC*) direction with respect to boresight; (*ii*) $D(\theta,\phi) \leq D(\theta_0 + \theta^{EoC}, \phi) - 20$ [dB] (i.e., $SLL(\theta,\phi) \leq -20$ [dB], being $SLL(\theta,\phi)$ the *SLL* evaluated with respect to the *EoC* angle) when $(\theta_0 + \theta_1) \leq \theta \leq (\theta_0 + \theta_2)$, $\theta_1$ and $\theta_2$ being the *EoC* angle of the nearest "iso-color" beam [22] and the (maximum) inside Earth angle [22]; (*iii*) $D(\theta,\phi) \leq D(\theta_0 + \theta^{EoC}, \phi) - 10$ [dB] (i.e., $SLL(\theta,\phi) \leq -10$ [dB]) for $\theta \geq \theta_2$. Towards this end, the *MS-OTM* process has been carried out by customizing (5) so that

$$\Phi(\mathbf{c}) \triangleq \max_{(\theta_0,\phi)\in\zeta} \{SLL[\mathbf{P}(\theta_0,\phi;\mathbf{c})]\} \qquad (13)$$

where $\zeta \triangleq \{0 \leq \theta_0 \leq 1.12$ [deg]; $0 \leq \phi \leq 360$ [deg]$\}$.

The array has been assumed to be arranged on a lattice of $N \times M = 78 \times 78$ positions spaced by $d_x = d_y = 0.84 \lambda$ (i.e., an aperture $R$ of dimensions $64.68 \lambda \times 64.68 \lambda$ having maximum diagonal dimension of about $91.5 \lambda$, then smaller than $120 \lambda$), while square open-ended waveguides of size $0.82 \lambda$ along the $x$ and $y$ axis, whose embedded element pattern - when considering the electromagnetic interactions of two rings of neighboring elements [Fig. 14(*b*)] - is shown in Fig. 14(*a*), have been chosen as radiating elements. By changing different values of $L$ and $S$ and setting the *IGA* population size to $P = 3 \times V$, the best trade-off (requirement-fitting vs. *MSTA*-complexity, $Q$) solution synthesized by the *MS-OTM* turned out to be that with $S = 3$ and $L = 9$ ($\rightarrow \hat{L} = 3$) shown in Fig. 15(*a*). Such a *MSTA* contains $Q = 372$ tiles ($Q_L = 38$ and $Q_S = 334$), which means a saving of 93.89 % *TRM*s with respect to a *FPA*. The corresponding power pattern when the beam is steered at ($\theta_0$, $\phi_0$) = (0, 0) [deg] and its principal cuts are given in Fig. 15(*b*) and Figs. 15(*c*)-(*d*), respectively. Moreover, the values of the sub-array phases [Figs. 16(*a*)-(*b*)] and the pattern cuts [Figs. 16(*c*)-(*d*)] when steering the main lobe towards ($\theta_0$, $\phi_0$) = (1.12, 0.0) [deg] [Fig. 16(*a*) and Fig. 16(*c*)] and ($\theta_0$, $\phi_0$) = (1.12, 90.0) [deg] [Fig. 16(*b*) and Fig. 16(*d*)] are reported, as well. Besides the pictorial proofs of the fitting of the project requirements [Figs. 15(*c*)-15(*d*); Figs. 16(*c*)-16(*d*)], the reliability of the synthesized *MSTA* is also quantitatively confirmed by the numerical values of the pattern features. As a matter of fact, both the maximum value of $SLL(\theta_0, \phi)$ and the minimum value of $D(\theta_0 + \theta^{EoC}, \phi)$ within the angular region $\zeta$ fulfill the project constraints (i.e., $\max_{(\theta_0,\phi)\in\zeta}\{SLL(\theta_0, \phi)\} = -20.13$ [dB] and $\min_{(\theta_0,\phi)\in\zeta}\{D(\theta_0 + \theta^{EoC}, \phi)\} \ge 43.85$ [dBi]). As for the maximum directivity loss with respect to the broadside case, it turns out to be $D_{loss} = 0.14$ [dBi], being $D = 46.32$ [dBi] the peak directivity of the broadside directed beam and $D = 46.18$ [dBi] the peak directivity of the beam directed towards ($\theta_0$, $\phi_0$) = (1.12, 0.0) [deg].

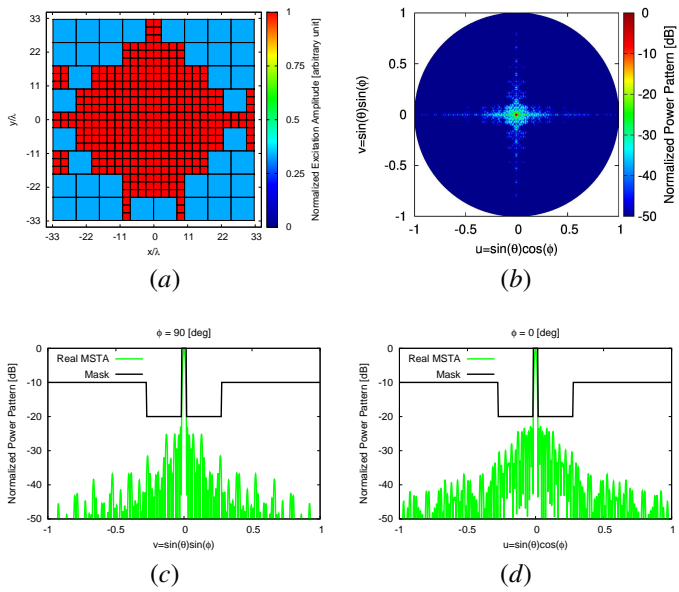

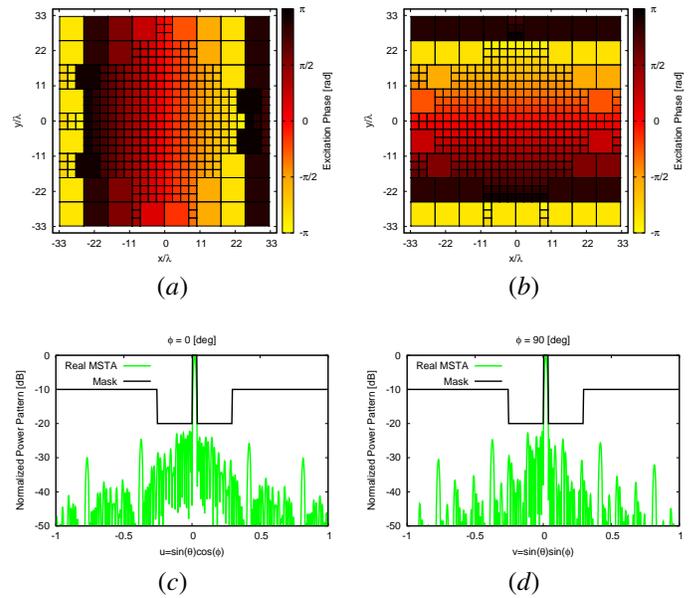

Figure 15.   *Numerical Assessment* (*Real MSTA*; $M = 78$, $N = 78$, $d = 0.84\lambda$; $S = 3$, $L = 9$; $Q = 372$: $Q_L = 38$, $Q_S = 334$) - Plots of (*a*) the equivalent element-level amplitude distribution and of the normalized power pattern (*b*) in the whole angular range ($-1 \le u \le 1$, $-1 \le v \le 1$) and along the $\phi = 90$ [deg] and (*d*) the $\phi = 0$ [deg] planes when steering the beam along ($\theta_0$, $\phi_0$) = (0.0, 0.0) [deg].

Figure 16.   *Numerical Assessment* (*Real MSTA*; $M = 78$, $N = 78$, $d = 0.84\lambda$; $S = 3$, $L = 9$; $Q = 372$: $Q_L = 38$, $Q_S = 334$) - Plots of (*a*) (*b*) the equivalent element-level phase distribution and of the cuts of the normalized power pattern along (*c*) the $\phi = 0$ [deg] and (*d*) the $\phi = 90$ [deg] planes when steering the beam towards (*a*)(*c*) ($\theta_0$, $\phi_0$) = (1.12, 0.0) [deg] and (*b*)(*d*) ($\theta_0$, $\phi_0$) = (1.12, 90.0)[deg].

## V. CONCLUSIONS AND REMARKS

This work has addressed the synthesis of rectangular phased arrays tiled with two-sized square tiles, each one controlled with a single isophoric amplifier and a single phase shifter, fitting user-defined constraints on the radiation pattern. An innovative tile-size tapering technique has been introduced and exploited to control the *SLL* by optimizing the tiling configuration.

The following main methodological novelties with respect to the state-of-the-art methods have been proposed:

- the design of rectangular isophoric phased arrays composed of two square tiles of different sizes to yield an advanced control of the beam-pattern features through an amplitude *tile-size* tapering of the single-element excitations;
- the exploitation of mathematical theorems from the combinatorial theory to provide the conditions for the complete tileability of a rectangular aperture with tiles composed of two square clusters of elementary radiators;
- a suitable integer coding of the admissible tilings that allows a drastic reduction of the cardinality of the solution space;
- the compact representation of the solution space through an acyclic graph;
- the development of an innovative *IGA* for an efficient exploration of the solution space/graph to effectively deal with large arrays, as well.

Representative examples from a wide set of numerical experiments concerned with ideal as well as real radiating elements have been discussed. From such a numerical assessment, the following outcomes can be drawn:

- the *IGA*-based *MS-OTM* guarantees a high success rate to converge towards solutions equal or close to the global optimum ones;
- the proposed tile-size tapering technique and the related *MSTA* architecture turn out to be effective in the control of the pattern features to fit real-problem specifications and requirements such as those from space (sensing and communications) applications.

Future research activities, beyond the scope of the current work, will be aimed at improving the computational efficiency of the solution space sampling by exploiting combinatorial optimizer intrinsically designed for exploring graphs-modeled spaces [42]. Moreover, the whole synthesis framework will be extended to planar arrays having arbitrary aperture shapes, not fully tileable apertures, as well as multiple ($> 2$) square sizes to evaluate the trade-off between modularity-maintenance/manufacturing costs and array performance.

## Appendix I

The adjacency matrix of the solution graph $\mathcal{G}$ is the matrix $\mathbf{G} \triangleq \{g_{ij}; i, j = 1, ..., H\}$ of size $H \times H$ whose entries are equal to $g_{ij} = 1$ when the node $\boldsymbol{\alpha}_i$ is connected to the node $\boldsymbol{\alpha}_j$ and $g_{ij} = 0$, otherwise [38].

## Acknowledgements

A. Massa wishes to thank E. Vico for her never-ending inspiration, support, guidance, and help.